\documentclass{appolb}
\usepackage{epsfig}

\begin{document}

\vspace*{-10mm}
{\small\hspace*{\fill} WUE-ITP-99-026

\hspace*{\fill} hep-ph/9910513}

\vspace*{-12mm}
\title{Selectron Mass Effects in Neutralino Production%
}
\author{S. Hesselbach\footnote{e-mail: hesselb@physik.uni-wuerzburg.de}, 
  H. Fraas\thanks{e-mail: fraas@physik.uni-wuerzburg.de}
\address{Institut f\"ur Theoretische Physik, Universit\"at W\"urzburg,\\
  Am Hubland, D-97074 W\"urzburg, Germany}
}
\maketitle
\begin{abstract}
We study the possibility to measure the masses of the selectrons in
neutralino production at an $e^+ e^-$ linear collider with polarized
beams. The cross sections and polarization asymmetries of neutralinos
with gaugino character strongly depend on the masses of the exchanged
selectrons. If the usual GUT relations of the selectron masses in the
MSSM are relaxed large effects are possible especially in the
polarization asymmetries. These can be used to determine the masses of
both selectrons.
\end{abstract}
\PACS{14.80.Ly, 13.88.+e, 12.60.Jv}
  
\section{Introduction}

The search for supersymmetric particles and the determination of their
masses is one of the main goals of a future $e^+ e^-$ linear collider. 
Especially the use of polarized beams plays an important role for the
measurement of the parameters of the underlying supersymmetric model.

In our contribution we study the effects of the selectron masses on
the cross sections, polarization asymmetries and decay angular
distributions of neutralino production
in $e^+ e^-$ annihilation. We focus on the production of
light neutralinos $e^+ e^- \to \tilde{\chi}^0_1 \tilde{\chi}^0_2$ in
the Minimal Supersymmetric Standard Model (MSSM) \cite{haberkane} with
GUT relation for the gaugino mass parameters $M_1/M_2 =
5/3\tan^2\theta_W$. We assume that the masses of the left
selectron $m_{\tilde{e}_L}$ and of the right selectron
$m_{\tilde{e}_R}$ are both independent parameters of the model. The
motivation for this is that in models with new U(1) factors in the
gauge group also additional $D$-terms appear in the mass terms of the
selectrons \cite{mselE6}. 
These $D$-terms are of the order of the masses $m_{Z'} =
\mathcal{O}(1~\mathrm{TeV})$ of the corresponding
new gauge bosons which results in larger differences between
$m_{\tilde{e}_L}$ and $m_{\tilde{e}_R}$ and even $m_{\tilde{e}_R} >
m_{\tilde{e}_L}$ is possible. The character of the light
neutralinos in these extended supersymmetric models which can be 
motivated by superstring theory \cite{hr} is in most cases very
similar to the MSSM \cite{lcpaper,dis}.
Therefore in the following the selectron mass effects in neutralino
production will be discussed in the MSSM.

\section{Production of neutralinos}

In $e^+ e^-$ annihilation the higgsino components of the neutralinos
are produced only via $s$ channel exchange of $Z$ bosons whereas the
gaugino components are produced
only via $t$ and $u$ channel exchange of selectrons
\cite{bartlfraasneutprod}.
So we choose a scenario
\begin{equation} \label{scen}
M_2 = 200~\mathrm{GeV},\quad \mu = 350~\mathrm{GeV},\quad \tan\beta = 3
\end{equation}
of the neutralino parameters where $\tilde{\chi}^0_1$
($m_{\tilde{\chi}^0_1} = 93$ GeV) and $\tilde{\chi}^0_2$
($m_{\tilde{\chi}^0_2} = 175$ GeV) have both a large gaugino component
($96.5\;\%$ and $87.9\;\%$ respectively) to analyze the selectron mass
effects. Then the contribution of the $Z$ exchange can be neglected
and for longitudinally polarized beams
the total cross section \cite{bartlfraasneutprod, christova} consists of
two terms describing the exchange of left and right
selectrons, respectively $\sigma \approx \sigma_{\tilde{e}} = 
  \sigma_{\tilde{e}_L} + \sigma_{\tilde{e_R}}$.
The structure of $\sigma_{\tilde{e}_{L/R}}$ is
\begin{equation}
\sigma_{\tilde{e}_{L/R}} = (f^{L/R}_{e1} f^{L/R}_{e2})^2
  \left[(1- P_-P_+) \mp (P_- - P_+)\right] f(s,m_{\tilde{e}_{L/R}})
\end{equation}
with the $\tilde{\chi}^0_i \tilde{e}_{L/R} e_{L/R}$ coupling
$f^{L/R}_{ei}$ \cite{bartlfraasneutprod} and the polarization
$P_{-/+}$ of the electron and 
positron beam, respectively. Thus the selectron mass effects in the cross
section $\sigma$ and in the polarization asymmetry for electron beam
polarization of $90\;\%$
\begin{equation}
  A_\mathrm{LR} = \frac{\sigma(P_- = -0.9) - \sigma(P_- =
  +0.9)}{\sigma(P_- = -0.9) + \sigma(P_- = +0.9)} 
\end{equation}
depend on the ratio 
\begin{equation}
  \label{rf}
  r_f = (f^R_{e1} f^R_{e2})^2/(f^L_{e1} f^L_{e2})^2 \, .
\end{equation}
In scenario (\ref{scen}) with $r_f = 0.19$ the exchange of
$\tilde{e}_L$ dominates.

\begin{figure}
\centering
\begin{picture}(12.6,9.15)

  \put(0,5){\epsfig{file=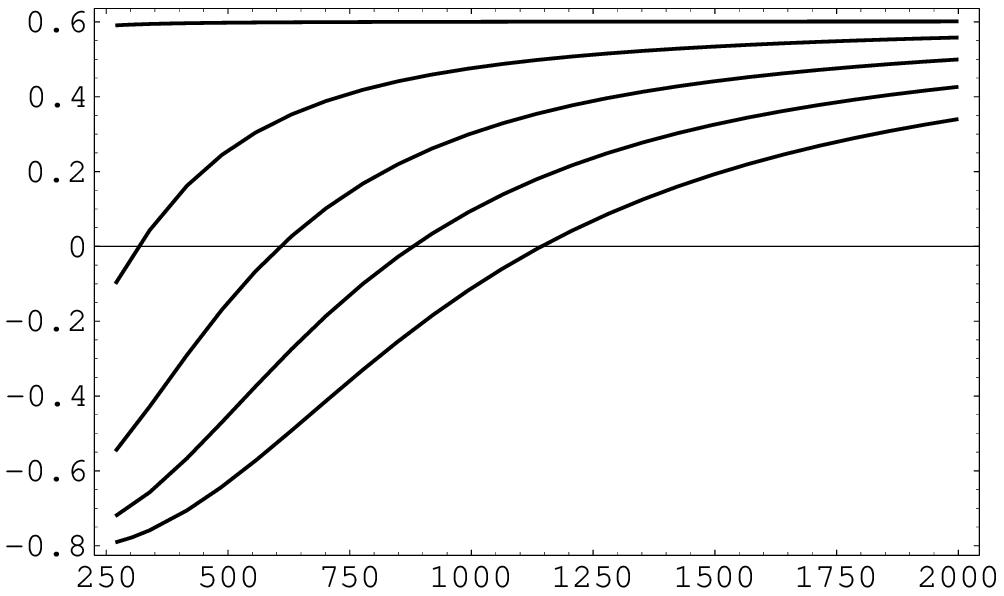,scale=.6}}
  \put(4.8,4.7){\footnotesize$\sqrt{s}/$GeV}
  \put(.05,8.75){\footnotesize$A_\mathrm{LR}$}
  \put(4,5.5){\footnotesize $m_{\tilde{e}_L}/$GeV}
  \put(1.7,5.5){\footnotesize 1000}
  \put(.7,8.2){\footnotesize 200}
  \put(3,8.8){\small (a)}

  \put(6.5,5){\epsfig{file=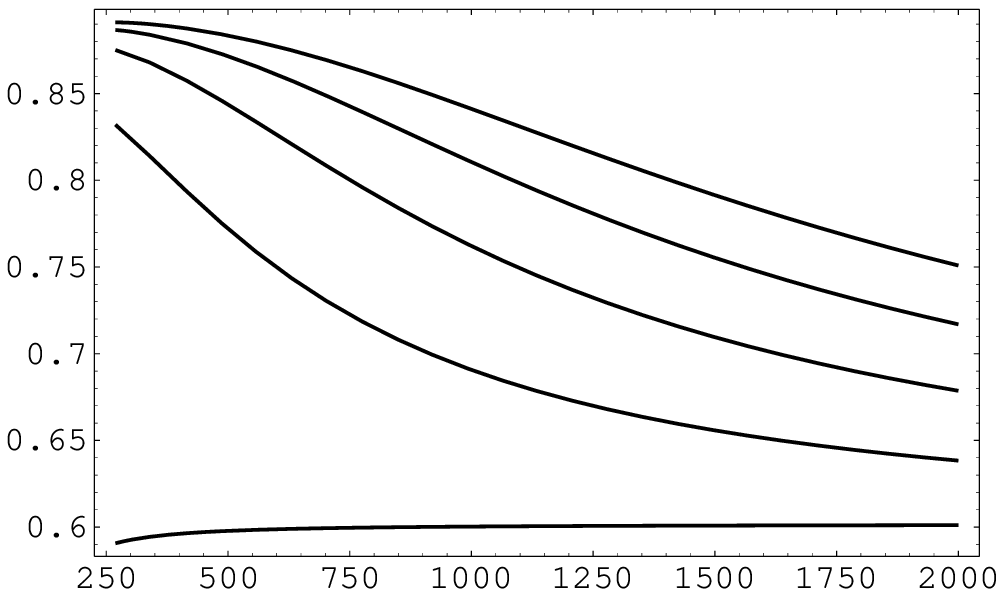,scale=.6}}
  \put(11.3,4.7){\footnotesize$\sqrt{s}/$GeV}
  \put(6.55,8.75){\footnotesize$A_\mathrm{LR}$}
  \put(10.5,8.2){\footnotesize $m_{\tilde{e}_R}/$GeV}
  \put(9.3,8.2){\footnotesize 1000}
  \put(7.3,5.5){\footnotesize 200}
  \put(9.5,8.8){\small (a)}

  \put(.2,.4){\epsfig{file=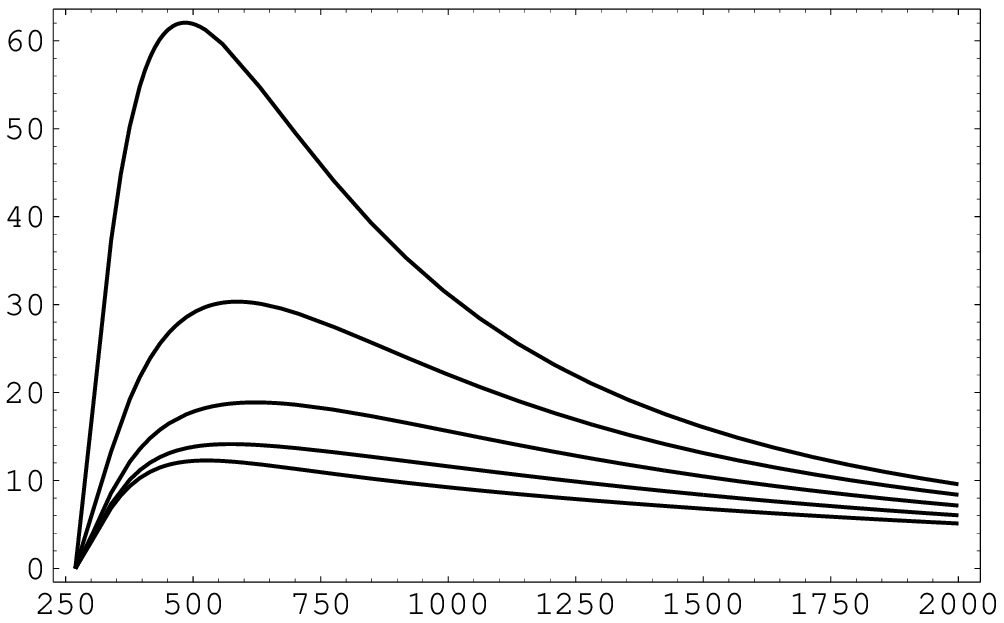,scale=.58}}
  \put(4.8,.1){\footnotesize$\sqrt{s}/$GeV}
  \put(.2,4.15){\footnotesize$\sigma/$fb}
  \put(3.9,3.6){\footnotesize $m_{\tilde{e}_L}/$GeV}
  \put(1.9,3.6){\footnotesize 200}
  \put(1.1,.95){\footnotesize 1000}
  \put(3,4.2){\small (b)}

  \put(6.6,.4){\epsfig{file=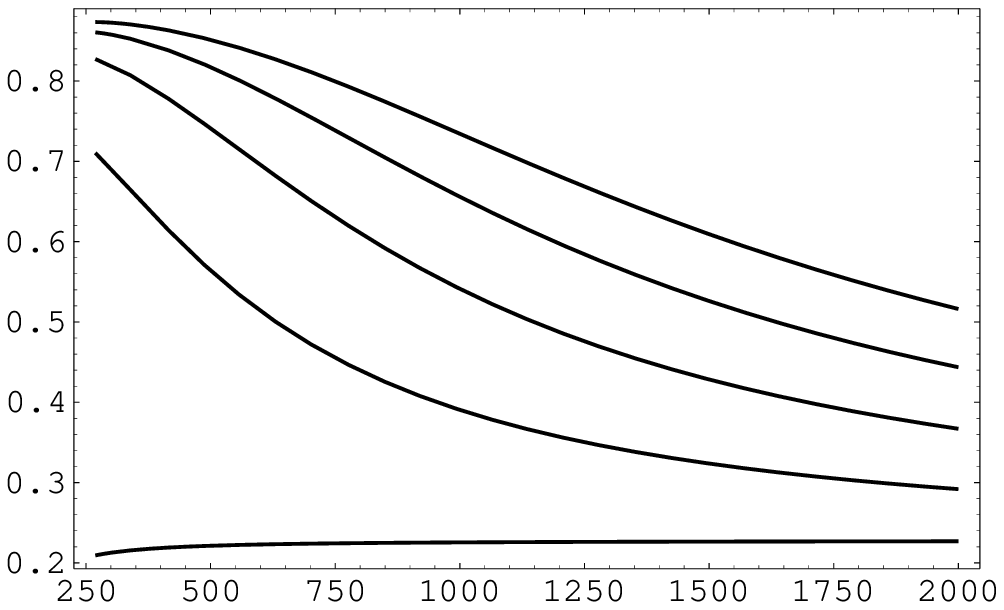,scale=.59}}
  \put(11.3,.1){\footnotesize$\sqrt{s}/$GeV}
  \put(6.55,4.15){\footnotesize$A_\mathrm{LR}$}
  \put(10.5,3.6){\footnotesize $m_{\tilde{e}_R}/$GeV}
  \put(9,3.6){\footnotesize 1000}
  \put(7.3,.95){\footnotesize 200}
  \put(9.5,4.2){\small (b)}
\end{picture}

\begin{minipage}[t]{6cm}
\caption{\label{fig1}
  (a) Polarization asymmetries $A_\mathrm{LR}$ and (b) cross
  sections $\sigma$ of the process $e^+ e^- \to 
  \tilde{\chi}^0_1 \tilde{\chi}^0_2$ for $m_{\tilde{e}_R} = 200$ GeV,
  $m_{\tilde{e}_L} = 200$, 400, 600, 800 and 1000~GeV and $P_+ = 0$.}
\end{minipage}
\hfill
\begin{minipage}[t]{6cm}
\caption{\label{fig2}
  Polarization asymmetries $A_\mathrm{LR}$ of the process $e^+ e^- \to
  \tilde{\chi}^0_1 \tilde{\chi}^0_2$ for $m_{\tilde{e}_L} = 200$ GeV,
  $m_{\tilde{e}_R} = 200$, 400, 600, 800 and 1000~GeV and (a) $P_+ =
  0$, (b) $P_+ = -50$\,\%.}
\end{minipage}
\end{figure}

Figs.~\ref{fig1}(a), (b) show $A_\mathrm{LR}$ and $\sigma$ for different
values of $m_{\tilde{e}_L}$ and a fixed value of
$m_{\tilde{e}_R}$. Because of the dominating $\tilde{e}_L$ exchange
$\sigma$ drops significantly with increasing $m_{\tilde{e}_L}$. 
Especially near threshold also
$A_\mathrm{LR}$ strongly depends on $m_{\tilde{e}_L}$. For
$m_{\tilde{e}_L} = m_{\tilde{e}_R}$ it is approximately independent of
energy
$A_\mathrm{LR} \approx 0.9 (1 - r_f)/(1 + r_f) \approx +60\;\%$.
For $m_{\tilde{e}_L}
\gg m_{\tilde{e}_R}$ it tends to the value $-90\;\%$ at threshold,
whereas for increasing energy it runs asymptotically to the value for
equal masses. So $A_\mathrm{LR}$ has different sign near threshold
depending on $m_{\tilde{e}_L}$.

In Fig.~\ref{fig2}(a) $A_\mathrm{LR}$ is depicted for fixed
$m_{\tilde{e}_L}$ and different $m_{\tilde{e}_R}$. 
For $m_{\tilde{e}_R} \gg m_{\tilde{e}_L}$ it tends to
the value $+90\;\%$ at threshold. 
Since the dominating \nopagebreak
contribution comes from $\tilde{e}_L$ exchange
the dependence on $m_{\tilde{e}_R}$ is weaker than \pagebreak
that on
$m_{\tilde{e}_L}$ in Fig.~\ref{fig1}(a).
For fixed $m_{\tilde{e}_L}$ and increasing $m_{\tilde{e}_R}$
also $\sigma$ drops less than in Fig.~\ref{fig1}(b)
(\eg $\sigma = 52$~fb for $m_{\tilde{e}_L}=200$~GeV, $m_{\tilde{e}_R}
= 1000$~GeV and
$\sqrt{s} = 500$~GeV). The dependence on $m_{\tilde{e}_R}$ can be enlarged if
additionally the positron beam is polarized (Fig.~\ref{fig2}(b)).

\begin{figure}
\begin{picture}(12.6,6.6)

  \put(0,.4){\epsfig{file=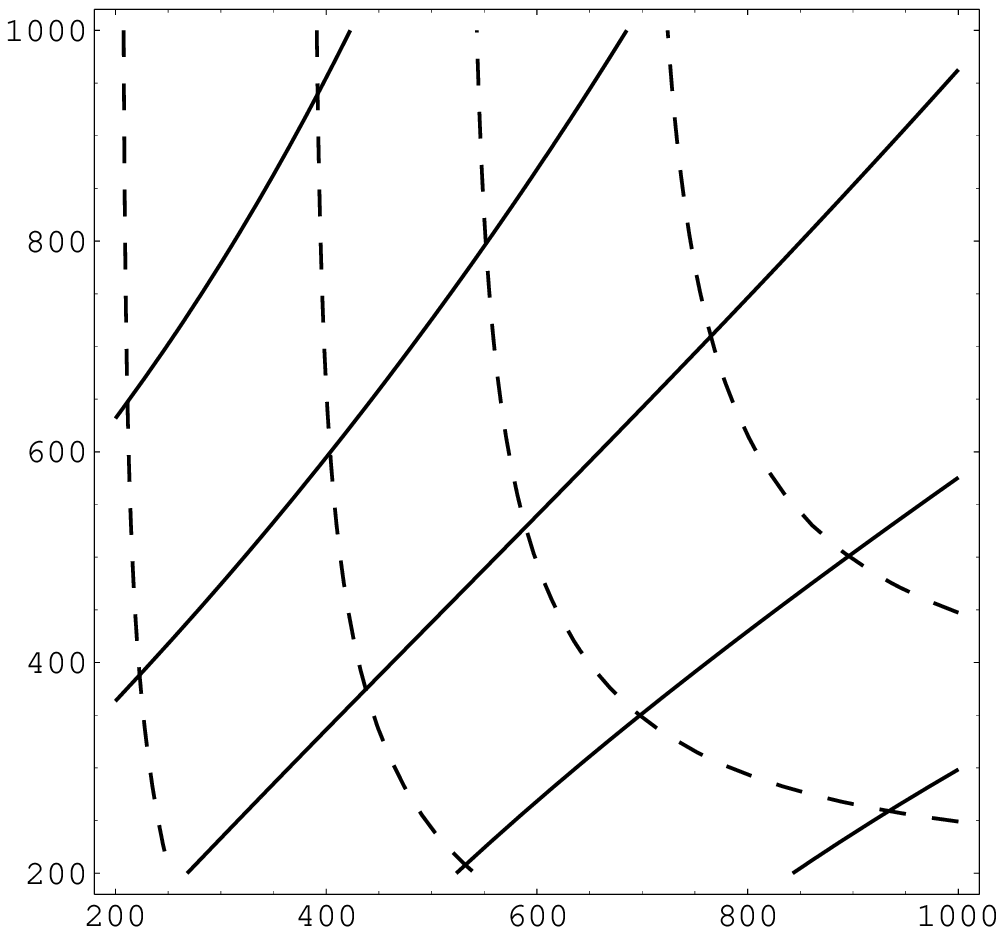,scale=.61}}
  \put(4.7,.1){\footnotesize $m_{\tilde{e}_L}/$GeV}
  \put(0,6.35){\footnotesize $m_{\tilde{e}_R}/$GeV}
  \put(.85,5.85){\scriptsize 50} \put(1.6,5.85){\scriptsize 20}
  \put(3.05,5,85){\scriptsize 10} \put(4.2,5,85){\scriptsize 5}
  \put(.9,3.65){\scriptsize $+0.85$} \put(1,1.95){\scriptsize $+0.75$}
  \put(1.4,.85){\scriptsize $+0.5$} \put(3.2,.85){\scriptsize 0}
  \put(4.2,.85){\scriptsize $-0.5$}

  \put(6.55,.4){\epsfig{file=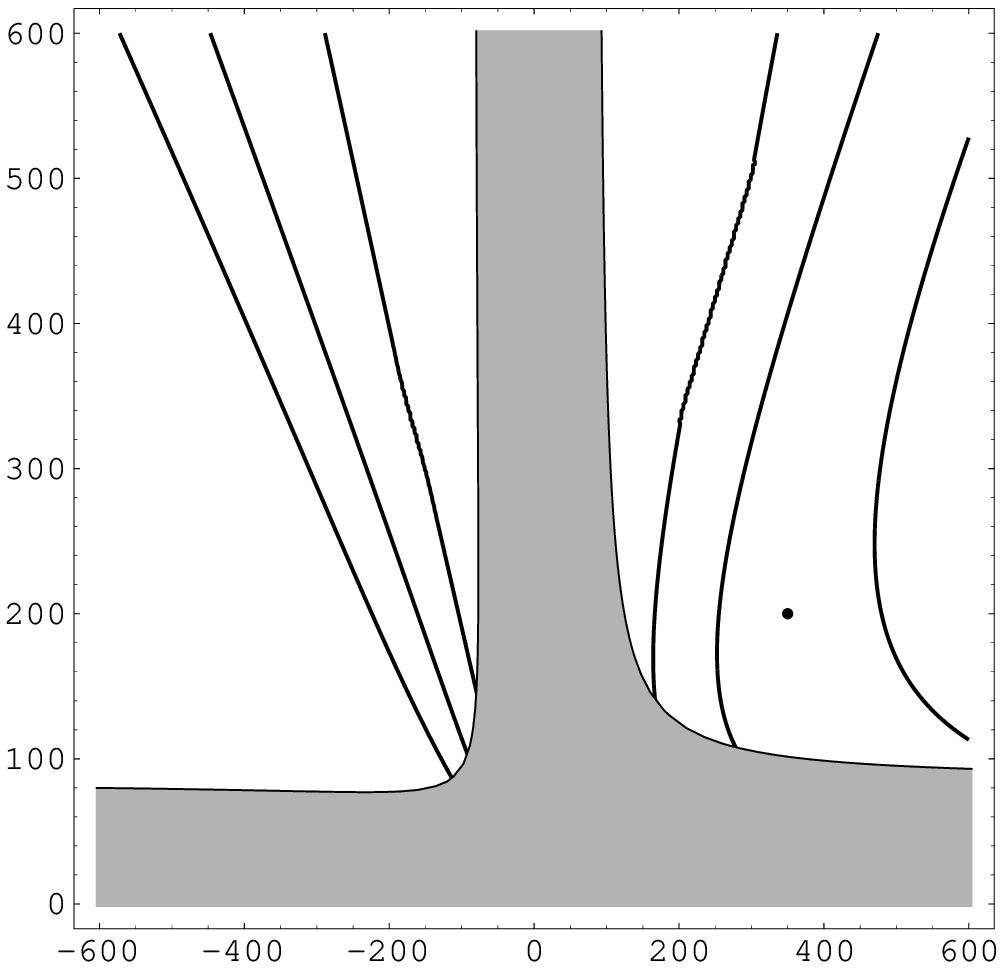,scale=.59}}
  \put(9.3,.1){\footnotesize $\mu/$GeV}
  \put(6.55,6.35){\footnotesize $M_2/$GeV}
  \put(7.15,4.65){\scriptsize $0.05$}
  \put(8,5.85){\scriptsize 1} \put(8.65,5.85){\scriptsize 20} 
  \put(10.8,5.85){\scriptsize 20} \put(11.5,5.85){\scriptsize 1}
  \put(11.9,5.5){\scriptsize $0.05$}
  \put(11,2.3){\scriptsize scen.}
\end{picture}
\begin{minipage}[t]{6.1cm}
\caption{\label{fig3}%
Contours of the polarization asymmetry $A_\mathrm{LR}$ for
  $P_+ = 0$ (solid) and of the cross section (in fb) for $P_+ = P_- =
  0$ (dashed) of the process $e^+ e^- \to \tilde{\chi}^0_1
  \tilde{\chi}^0_2$ for $\sqrt{s} = 500$ GeV.}
\end{minipage}
\hfill
\begin{minipage}[t]{5.9cm}
\caption{\label{fig4}%
Contours of the ratio $r_f$ (\ref{rf}) for $\tan\beta = 3$. Also
  shown are the \mbox{experimentally} excluded parameter space
  (shaded) and the scenario (\ref{scen}).}
\end{minipage}
\end{figure}

Fig.~\ref{fig3} shows the contours of $A_\mathrm{LR}$ and $\sigma$ in
the $m_{\tilde{e}_L}$-$m_{\tilde{e}_R}$ parameter space. These two
observables form a network which in principle allows to determine the two
selectron masses, if the parameters of the neutralino sector are known.
In Fig.~\ref{fig4} the contours of $r_f$ are plotted
in the $M_2$-$\mu$ parameter space. Above the lines $r_f = 20$ the
higgsino character of the neutralinos dominates, so the selectron mass
effects are not visible. In the lower left corner (large negative
$\mu$) $r_f$ is very small. Then only the left selectrons
contribute to the cross section and therefore $A_\mathrm{LR} \approx
+90\;\%$ independently of $m_{\tilde{e}_R}$. But in large regions of
the parameter space not too far below the lines $r_f = 0.05$
both selectrons contribute and the selectron mass effects in the
neutralino production are measurable.

\begin{figure}
\centering
\begin{picture}(12.6,4.5)

  \put(0,.4){\epsfig{file=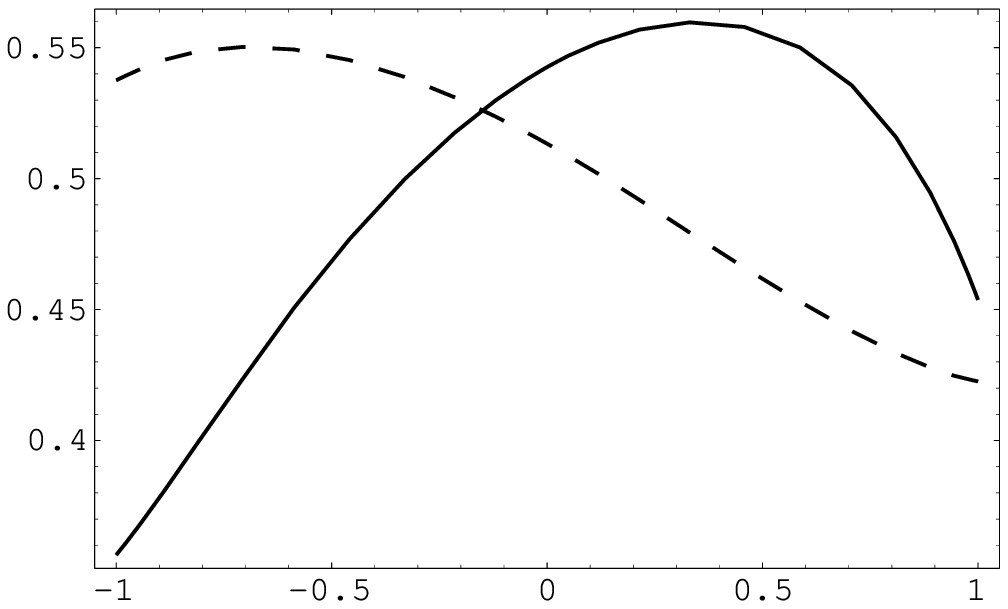,scale=.592}}
  \put(0,4.25){\footnotesize
 $\frac{\mathrm{d}\tilde{\sigma}}{\mathrm{d}\cos\theta_-}/\tilde{\sigma}$}
  \put(3,.1){\footnotesize $\cos\theta_-$}
  \put(1,.9){\scriptsize $\tilde{\sigma} = 17.1$~fb, 
    $A_\mathrm{FB} = +7.6\;\%$}
  \put(2.5,1.5){\scriptsize $\tilde{\sigma} = 1.4$~fb, 
    $A_\mathrm{FB} = -7.6\;\%$}

  \put(6.4,.4){\epsfig{file=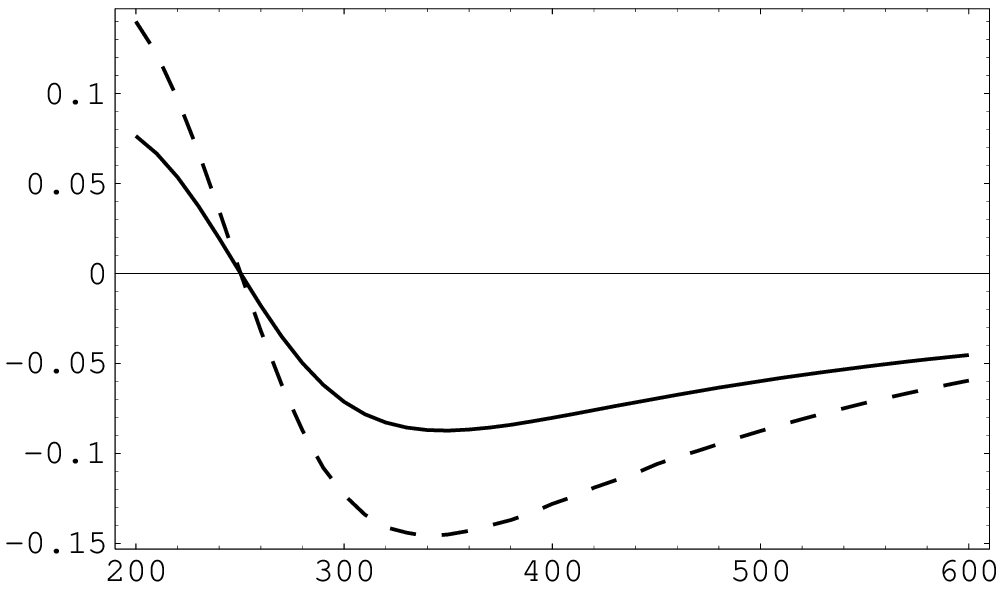,scale=.61}}
  \put(6.5,4.2){\footnotesize $A_\mathrm{FB}$}
  \put(11.2,.1){\footnotesize $m_{\tilde{e}_L}/$GeV}

\end{picture}

\begin{minipage}[t]{6cm}
\caption{\label{fig5}
  Decay angular distributions of the decay leptons
  $\frac{\mathrm{d}\tilde{\sigma}}{\mathrm{d}\cos\theta_-}/\tilde{\sigma}$
  for $P_- = -90\;\%$, $P_+ = 0$, $\sqrt{s} = 500$ GeV,
  $m_{\tilde{e}_R} = 200$ GeV and $m_{\tilde{e}_L}= 200$ GeV (solid)
  or $m_{\tilde{e}_L}= 400$ GeV (dashed).}
\end{minipage}
\hfill
\begin{minipage}[t]{6cm}
\caption{\label{fig6}
  Forward-backward asymmetries $A_\mathrm{FB}$ of the decay leptons
  for $P_- = -90\;\%$, $P_+ = 0$, $m_{\tilde{e}_R} = 200$ GeV and
  $\sqrt{s} = 500$ GeV (solid) or $\sqrt{s} = 300$ GeV (dashed).}
\end{minipage}
\end{figure}

In Fig.~\ref{fig5} the decay angular distributions
$\frac{\mathrm{d}\tilde{\sigma}}{\mathrm{d}\cos\theta_-}$ for the
leptonic decay of the $\tilde{\chi}^0_2$ are shown for two values of
$m_{\tilde{e}_L}$. They are computed with full spin correlations
between production and decay \cite{gudi,sitges} and are normalized on the
total cross section $\tilde{\sigma}$ of the combined process $e^+ e^-
\to \tilde{\chi}^0_1 \tilde{\chi}^0_2 \to \tilde{\chi}^0_1
\tilde{\chi}^0_1 e^+ e^-$, where $\theta_-$ is the angle between
the ingoing $e^-$ and the outgoing $e^-$. The angular distributions
also show a strong dependence on the selectron masses for the chosen
beam polarization. The forward-backward asymmetries $A_\mathrm{FB}$
\cite{gudi} change sign with increasing $m_{\tilde{e}_L}$
(Fig.~\ref{fig6}). The dependence of $A_\mathrm{FB}$ on
$m_{\tilde{e}_L}$ is larger for $\sqrt{s}$ closer to threshold.

\section{Conclusions and outlook}

The cross section, polarization asymmetry and forward-backward
asymmetry of neutralino production strongly depend
on the selectron masses for suitable beam polarization
especially near threshold.
A simultaneous measurement of $\sigma$ and $A_\mathrm{LR}$ allows 
in principle the
determination of the selectron masses. This also provides a
possibility to discriminate between the MSSM and extended models.
Thus beam polarization is an important feature of a future linear collider
to determine the parameters of the underlying model.

We are grateful to G. Moortgat-Pick for many valuable discussions and for
providing the program to compute the angular distributions.
S. H. thanks H. Czy\.{z} and the other organizers of the XXIII School
of Theoretical Physics for the friendly atmosphere during the
conference. 
This work was supported by the Bundesministerium f\"ur
Bildung und Forschung (BMBF) under contract No. 05 HT9WWA 9, by the
Stiftung f\"ur deutsch-polnische Zusammenarbeit, Warszawa and by
the Fonds zur F\"orderung der wissenschaftlichen Forschung of Austria,
project No.\ P13139-PHY.


\begin{thebibliography}{9}

\bibitem{haberkane}H.E. Haber, G.L. Kane, Phys. Rep. \textbf{117}
  (1985) 75. 

\bibitem{mselE6}M. Drees, Nucl. Phys. \textbf{B 298} (1988) 333;
  H.-C. Cheng, L.J. Hall, Phys. Rev. \textbf{D 51} (1995) 5289;
  C. Kolda, S.P. Martin, Phys. Rev. \textbf{D 53} (1996) 3871.

\bibitem{hr} J.L. Hewett, T.G. Rizzo, Phys. Rep. \textbf{183} (1989)
  193.

\bibitem{lcpaper}S. Hesselbach, F. Franke, H. Fraas,  
  in \emph{$e^+e^-$ Linear Colliders: Physics and Detector Studies,
  Part E}, Contributions to the Workshops -- Frascati, London, Munich,
  Hamburg, Ed. R. Settles
  (\mbox{DESY~97-123E}, Hamburg, 1997) p. 479.

\bibitem{dis}S. Hesselbach, Ph.D. thesis, University of W\"urzburg, 1999.

\bibitem{bartlfraasneutprod}A. Bartl, H. Fraas, W. Majerotto,
  Nucl. Phys. \textbf{B 278} (1986) 1.

\bibitem{christova}E.C. Christova, N.P. Nedelcheva,
  Int. J. Mod. Phys. \textbf{A 5} (1990) 2241.

\bibitem{gudi}G. Moortgat-Pick, H. Fraas, A. Bartl, W. Majerotto,
  Acta Phys. Polon. \textbf{B 29} (1998) 1497;
  Eur. Phys. J. \textbf{C 9} (1999) 521; Eur. Phys. J. \textbf{C 9} 
  (1999) 549(E).

\bibitem{sitges}G. Moortgat-Pick, S. Hesselbach, F. Franke, H. Fraas, 
  WUE-ITP-99-023, hep-ph/9909549.

\end{thebibliography}
\end{document}